\documentclass[aps,prl,twocolumn,groupedaddress,showpacs]{revtex4}
\usepackage[pdftex]{graphicx}   % need for figures
\usepackage{nicefrac}
\usepackage{amsmath}
\usepackage{amssymb}
\usepackage{color}
\begin{document}
\hyphenation{AuGe PHOIBOS SAES}
\title{Structural Examination of Au/Ge(001) by Surface X-Ray Diffraction and Scanning Tunneling Microscopy}
\author{S.~Meyer$^{1}$}
\author{T.~Umbach$^{1}$}
\author{C.~Blumenstein$^{1}$}
\author{J.~Sch\"afer$^{1}$}
\author{R.~Claessen$^{1}$}
\author{S.~Sauer$^{2}$}
\author{S.J.~Leake$^{3}$}
\author{P.R.~Willmott$^{3}$}
\author{M.~Fiedler$^{4}$}
\author{F.~Bechstedt$^{4}$}
\affiliation{$^{1}$Experimentelle Physik 4 and R\"ontgen Research Center for Complex
Materials, Julius-Maximilians-Universit\"at W\"urzburg, D-97074 W\"urzburg, Germany}
\affiliation{$^{2}$Institut f\"ur Physik, Albert-Ludwigs-Universit\"at Freiburg, D-79104 Freiburg, Germany}
\affiliation{$^{3}$Swiss Light Source, Paul Scherrer Institut, CH-5232 Villigen, Switzerland}
\affiliation{$^{4}$Institut f\"ur Festk\"orpertheorie und -optik, Friedrich-Schiller-Universit\"at, D-07743 Jena, Germany}
\date{\today}

\begin{abstract}
The one-dimensional reconstruction of Au/Ge(001) was investigated by means of autocorrelation functions from surface x-ray diffraction (SXRD) and scanning tunneling microscopy (STM). Interatomic distances found in the SXRD-Patterson map are substantiated by results from STM. The Au coverage, recently determined to be \nicefrac{3}{4} of a monolayer of gold, together with SXRD leads to three non-equivalent positions for Au within the c(8$\times$2) unit cell. Combined with structural information from STM topography and line profiling, two building blocks are identified: Au-Ge hetero-dimers within the top wire architecture and Au homo-dimers within the trenches. The incorporation of both components is discussed using density functional theory and model based Patterson maps by substituting Germanium atoms of the reconstructed Ge(001) surface.
\end{abstract}

%\pacs{71.20.-b,71.27.+a,71.30.+h,79.60.-i}

\maketitle
The pristine germanium (001)-surface exhibits two reconstructions at room temperature, a static c(4$\times$2) arrangement of buckled dimers forming one-dimensional rows and its dynamic (2$\times$1)-counterpart consisting of flipping dimers \cite{Zandvliet2004457}. This surface acts as a template for the growth of one-dimensional (1D) chains after adsorption of metal atoms at elevated temperatures \cite{ISI:000272703300010}. Here, the reconstruction of Au on Ge(001) was found by STM to form an almost ideal 1D architecture where the chains are restricted to nearly single-atom width \cite{PhysRevLett.101.236802}. The electronic states close to the chemical potential are of 1D character \cite{PhysRevB.83.121411,PhysRevB.84.115411} with some indications of a lesser anisotropy at higher binding energies \cite{PhysRevB.84.115411,PhysRevB.80.081406}. The question whether the conduction channel is parallel \cite{PhysRevB.83.121411} or perpendicular to the wire direction \cite{PhysRevB.84.115411} is still unresolved. The latter case would lead to discontinuities within the wires in STM at low bias which was not observed \cite{PhysRevLett.101.236802,PhysRevB.78.233410}. Therefore the parallel scenario seems more favorable. Close inspection of the electronic states at the chemical potential yields a deviation from a common Fermi-liquid picture of 3D metals. Instead, Luttinger liquid behavior was observed in terms of a power-law scaling over energy and temperature of the density of states in the vicinity of the chemical potential \cite{DOI:10.1038/NPHYS2051}.
\par
Despite the extensive studies of the electronic properties of Au/Ge(001) the detailed structural atomic arrangement of this chain system remains unknown. The Au coverage as concluded from experiments ranges from $0.25$ to $1.2$~monolayer (ML) \cite{PhysRevB.78.233410,PhysRevB.70.233312,PhysRevB.80.081406}, while a recent study reports \nicefrac{3}{4} of a ML, accurately deduced from a calibrated sample and Auger electron spectroscopy \cite{PhysRevB.83.033302}. First STM data by Wang \textit{et al.} were interpreted in a double row scenario of Au-Au-dimers and mixed Au-Ge-dimers lying perpendicular to the wire direction but at the exact same height \cite{PhysRevB.70.233312,Wang2005126}. A second model also deduced from STM suggests buckled Ge dimers on top of the wires, and side walls consisting of gold $\sqrt{3}\times\sqrt{3}$-facets \cite{PhysRevB.78.233410,PhysRevB.80.081406}, although density functional theory (DFT) calculations predict such a model to be energetically unfavorable \cite{PhysRevB.81.075412}. Further STM results at 77 K resolved pronounced charge concentrations of V- and W-shape \cite{PhysRevB.83.035311}, also contradicting such facets. A recent temperature-dependent STM study could relate these shapes to the observed p(4$\times$1) superstructure spots in low-energy electron diffraction (LEED), which show a reversible, second order type phase transition at the critical temperature $T_{C}\sim 585$ K \cite{PhysRevLett.107.165702}. 

All of the previous reports on the atomic structure are based on STM. Yet this technique suffers from the limitation that the signal depends on the local density of states (LDOS) rather than topography. To circumvent this problem, we present a combined study of SXRD, STM and DFT calculations to obtain an insight into the atomic arrangement of the Au/Ge(001) chains. The corresponding autocorrelation function (Patterson map) from in-plane scattering data contains eight inequivalent vectors within the c(8$\times$2) unit cell. High-resolution STM images are used not only as a guide as to how to embed the Au atoms for a starting model, but also to calculate an autocorrelation map to cross-check the distances found in SXRD. Using the most accurately determined Au coverage of \nicefrac{3}{4} of a monolayer by Gallagher \textit{et al.} \cite{PhysRevB.83.033302} and associating the most pronounced Patterson-map peaks to Au-Au distances, yields two structural building blocks: single gold atoms embedded in the wires ridges and gold dimers located within the trenches. Both structural components are compatible with a model originally proposed by Sauer et al. \cite{PhysRevB.81.075412}, and exclude other proposed structures \cite{PhysRevB.78.233410,PhysRevB.80.081406}.

Sample preparation was performed on $0.4\ \Omega$cm commercial n-type (Sb doped) Ge(001) substrates. Wet chemical etching and oxidation were carried out \textit{ex-situ} to clean the sample before transfer into ultra-high vacuum (UHV) \cite{10.1063/1.3624902}. \textit{In-situ} preparation was performed in an UHV chamber with a base pressure of $1\times10^{-10}$~mbar. Transport to and x-ray diffraction at the MS Beamline X04SA, Swiss Light Source, Paul Scherrer Institut, Switzerland were carried out in a separate UHV transport chamber at a base pressure $< 1\times10^{-9}$ mbar. SXRD experiments were conducted at room-temperature with monochromatized $15$~keV x-rays. The diffractometer, housing a Pilatus 100k detector has an angular precision of $0.0025^\circ$, while the hexapod resolution is $0.0012^\circ$.

\begin{figure}[t!]
 \centering
		\includegraphics[width=\linewidth]{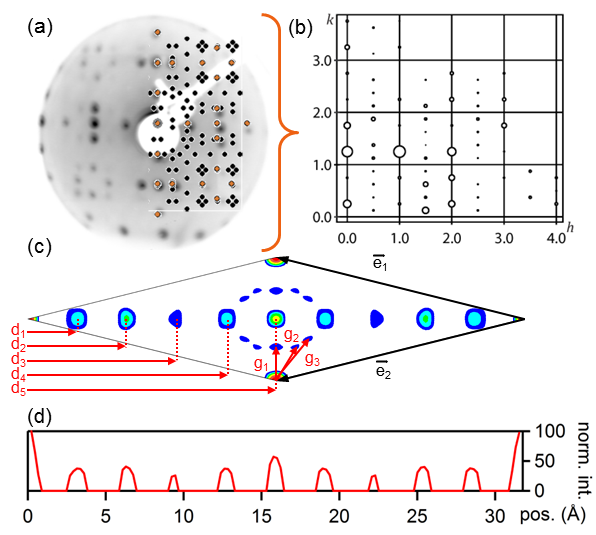}
		\caption[in-plane data and Patterson map]{(Color online)(a) LEED pattern at 22 eV. The basic c(8$\times$2) is indicated by the orange overlay, the p(4$\times$1) superstructure is shown in black. (b) SXRD in-plane scattering dataset for one domain only at $l=0.05\,$ r.l.u.; the diameter of the circles is proportional to the structure factor $\left|F_{hkl}\right|^2$. (c) Projected Patterson function of the electron density calculated from the fractional order in-plane reflections within the c($8 \times 2$) unit cell ($\vec{e}_{1/2}$). Vectors $\vec{d}_{1/2/3/4/5}$ are attributed to Au-Au distances, $\vec{g}_{1/2/3}$ to Au-Ge. (d) Line-profile across the central horizontal line of the Patterson map. Intensity is normalized to the peak at the point of origin (left corner of Patterson map).}
\label{fig1}
\end{figure}

Due to the Ge stacking sequence of ABAB, two equivalent surface domains exist, rotated by $90^\circ$ and separated by single atom height (1.4 \AA) steps. The Au/Ge(001) surface exhibits two reconstructions as illustrated from the corresponding low-energy electron pattern (22 eV) in Fig.\ref{fig1}(a). It contains a c(8$\times$2) reconstruction (orange dots in the overlay) representing the basic structure present at all temperatures and an additional superstructure of p(4$\times$1) type (black dots) which appears only below T$_C \sim$ 585 K \cite{PhysRevLett.107.165702}.

For the basic c(8$\times$2) reconstruction we have recorded $69$~symmetrically inequivalent reflexes in the $hk$-plane at lowest possible surface normal component $l=0.05$ r.l.u. (reciprocal lattice units) for the in-plane data set of Fig.~\ref{fig1}(b). Since both surface domains are equivalent and their diffraction spots do not overlap, it is sufficient to concentrate on one domain only. Thus, the presented in-plane dataset displays only even and odd values of 0.25 for $h$ and $k$, respectively with n $\in$ $\mathbb{N}_{0}$. Reflexes from the second domain are measured to check for consistency, but not shown here. The p(4$\times$1) superstructure, which is already very faint in LEED, was too weak to be addressed within the present SXRD data set.
Measured intensities are corrected for background, Lorentz-factor, polarization factor, and active sample area. Because SXRD only yields intensities, i.e. $|F_{hkl}|^2$, the phase information of the structure factor $F_{hkl}$ is lost, thereby excluding the direct calculation of the electron density by Fourier transformation. This problem is alleviated by the Patterson function $P(\vec{r})$:

\begin{eqnarray*}
P(\vec{r})&=& \int \rho(\vec{r'})\rho(\vec{r'}+\vec{r}) d\vec{r'}\\	
					&=&\sum_{hkl}\left|F_{hkl}\right|^2 exp(-i\vec{q}\vec{r}),
\end{eqnarray*}

\noindent
i.e., the autocorrelation function of the electron density $\rho(\vec{r})$ generated by applying a Fourier transformation directly to the corrected diffraction intensities, ignoring the unknown phases. All interatomic distances in $\rho(\vec{r})$ are also present in $P(\vec{r})$ allowing us to obtain lengths and directions, but not absolute positions. A contour plot of the Patterson function is presented in Fig.~\ref{fig1}(c). The point of origin for the Patterson map is at the left corner, also visible from the normalized intensities of the line-profile along the center of the unit cell of Fig.~\ref{fig1}(d) from the left to the right corner of the map. 

The intensity of any given peak in the Patterson function scales with the product of the atomic numbers of the contributing atoms. Hence, the most intense maxima in the line-profile are due to distances between Au atoms ($|\vec{d}_1|$ to $|\vec{d}_5|$). Not all of these vectors represent individual gold atoms, because this would result in a gold coverage of $\nicefrac{5}{4}$ ML. The most recent and accurate report quotes \nicefrac{3}{4} of a monolayer from a calibration sample and Auger spectroscopy \cite{PhysRevB.83.033302}. Thus, a c(8$\times$2) unit cell contains 6 gold atoms (3 per half unit cell) with 8 (4) Ge atoms underneath. 

The five vectors per half unit cell found in SXRD can be related to three gold atoms underneath by taking the 16 \AA\ periodicity of the wires into account. $\vec{d}_5$ can be generated by adding $\vec{d}_{1}+\vec{d}_{4}$ or $\vec{d}_{2}+\vec{d}_{3}$. Hence, $\vec{d}_{1}$ and $\vec{d}_{4}$ can match the same atom position (gold atom) if one vector starts at the corner of the unit cell and the other at the center (the same applies for $\vec{d}_{2}$ and $\vec{d}_{3}$). The remaining peaks (vectors $\vec{g}_{1/2/3}$ in Fig.~\ref{fig1}(c)) are attributed either to Au-Ge or Ge-Ge peaks. In a simple scattering picture these reflexes should decrease by a factor of 2.5 and 6 respectively according the ratio of their atomic mass numbers: $Z_{Au}^2$:$Z_{Au}\cdot Z_{Ge}$:$Z_{Ge}^2$=6:2.5:1, with $Z_{Au}$=79 and $Z_{Ge}$=32. From this simple argument the maxima of $\vec{g}_{1/2/3}$ are a factor of $\sim$ 2 lower in intensity and thus are attributed to Au-Ge distances. 

The Patterson map allows the exclusion of previously proposed complex structures, such as the Giant-Missing-Row model by van Houselt \textit{et al.} \cite{PhysRevB.78.233410}. Here, a $\sqrt{3}\times\sqrt{3}$-reconstruction of Au is built on (111)-facets of Ge as the sidewalls of the nanowires resulting in a Au coverage of 1 ML \cite{PhysRevB.81.075412}. To compensate the discrepancy in coverage, a modified version was proposed by Sauer et al. \cite{PhysRevB.81.075412} where initial Au-dimers on top switch to sidewall facets after relaxation in DFT. This Au-trimer stabilized germanium ridge model accounts for the correct coverage of \nicefrac{3}{4} of a ML, yet is incompatible with the Patterson map presented here, because there are insufficient atomic sites to accommodate all vectors found in SXRD.

\begin{figure}[t!]
	\centering
		\includegraphics[width=\linewidth]{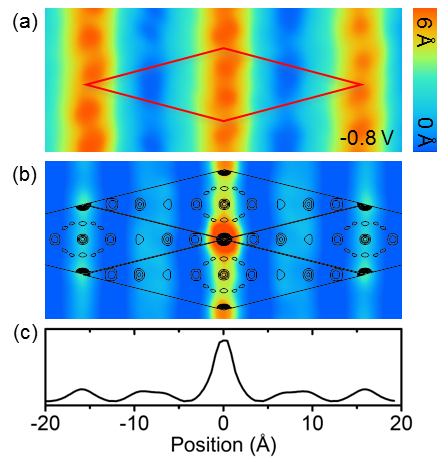}
	\caption[STM and its Autocorrelation]{(Color online)(a) Experimental STM image (occupied states at U = -0.8 V, I = 0.3 nA, T = 600 K)
exhibits a very faint zig-zag along the wire while the trenches are of low contrast. (b) Autocorrelation from STM showing sharp and broad intensity distributions. The overlay with the Patterson map from SXRD suggests good agreement of maxima coinciding on the shapes of intensity observed for($\vec{d}_{2/3/5}$). These are also found in horizontal line-profiles across the center maxima (c).}
\label{fig2}
\end{figure} 

Next, we address the question how the three Au atom per half unit cell can be arranged to account for the distances found in SXRD. Indications come from the STM topography in Fig.~\ref{fig2}(a), exhibiting a faint buckling along the wire, although with a small amplitude. The image was recorded at $\sim$600 K, where only the c(8$\times$2) reconstruction is present \cite{PhysRevLett.107.165702}. Previous line-profile analysis of the nanowire ridge are too sharp to be explained by a flat dimer \cite{PhysRevLett.101.236802,PhysRevLett.107.165702}. 
 
Two possible origins for the zig-zag appearance can be imagined: a structural buckling or electronic contrast originating from different orbitals. Both can be accounted for by a hetero-dimer of Au and Ge that may be buckled. Alternatively, a single atom whose neighboring atom along the chain is slightly shifted in-plane may mimic the observed zig-zag. For both options one of the three Au atoms per half unit cell is required. Experimental evidence for the buckled dimer was brought up recently by Mocking \textit{et al.}, who reported a dynamical flipping mode of wire segments close to defects in STM \cite{Mocking20102021}. 

With one of three atom of the Au coverage implemented in the wire ridge, the two remaining Au atoms must be located in the trenches. These appear homogeneous and flat in STM for all applied bias voltages, see also Fig.\ref{fig2}(a), leading to the conclusion of a flat Au homo-dimer as a structural building block. 

Before discussing the possibilities how to incorporate these two building blocks of homo- and hetero-dimers in the substrate, STM can be used to further verify the distances found in SXRD. For this purpose a autocorrelation map is generated from STM topography data ($\approx$60 nm)$^2$ containing only information on the basic c(8$\times$2) reconstruction (recorded above T$_C$). The unit cell shows two types of intensity profile, a sharp and a broad one, where the broad shape appears to be a result from a double row. Line-profiling across the central maxima in Fig.~\ref{fig2}(c) reveals a distance of 16 \AA\ ($\vec{d}_{5}$) between sharp lines, which is directly related to the wire separation. The distances from a sharp line to both maxima of a neighboring broad line are 7 \AA\ and 8.6 \AA, which can be related to $|\vec{d}_{2}|$ and $|\vec{d}_{3}|$ of the SXRD Patterson map within a 10\% error bar. Hence, the autocorrelation map from STM quantitatively matches three of the most intense maxima ($|\vec{d}|_{2/3/5}$) in SXRD, as indicated by the overlay of the corresponding Patterson map in Fig.~\ref{fig2}(b). The other distances found in SXRD might not be accessible by STM due to different contrast mechanism.

\begin{figure}[t!]
	\centering
		\includegraphics[width=\linewidth]{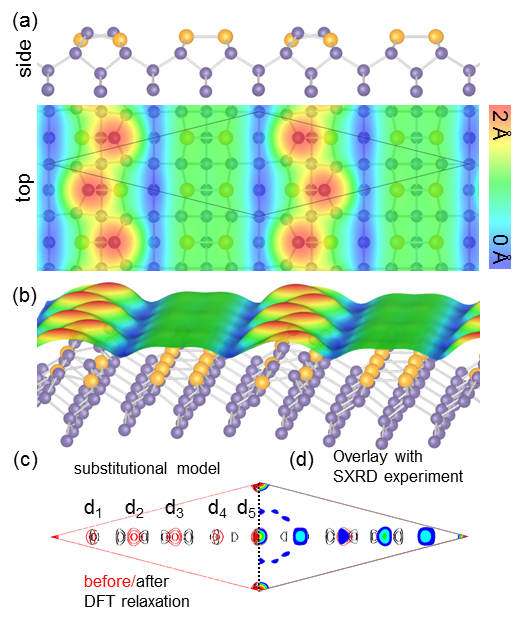}
		\caption[Substitutional model]{(Color online)(a) Cross-section and top view of relaxed DFT model (Au = yellow, Ge = purple-blue).
		Overlay showing calculated STM image for occupied states, U = -1 V. (b) Side view of (a). Pronounced charge in the wire direction originates from Ge, while the Au homo-dimers in the troughs are virtually featureless. (c) Calculated Patterson map for initial (red) and relaxed (black) model. Initial model reflects all main maxima of the experiment, see overlay in (d), but does not account for the correct intensity modulation. Main discrepancies arise after DFT relaxation for maxima corresponding to $\vec{d}_{2}$ and $\vec{d}_{3}$.}
	\label{fig3}
\end{figure}

Based on these considerations, a ``minimum structural model" is constructed with hetero- and homo-dimers as the main building blocks to account for the coverage of $\nicefrac{3}{4}$ of a ML of Au and the distances $|\vec{d}_1|$ to $|\vec{d}_5|$ found in SXRD and STM. The Au-Ge distances have to be neglected, to reduce the set of parameters. Both types of dimers can be implemented in the bare Ge(001) surface by substituting one (2$\times$1) double-row completely with Au homo-dimers and the neighboring row only partially, i.e. one of two Ge-dimer atoms is replaced by Au. Up to now no information on the vertical arrangement is implemented yielding the same height for both elements. This arrangement matches the suggestion of double-rows by Wang \textit{et al.} \cite{PhysRevB.70.233312}.

As a test for the ``minimum structural model", DFT calculations were performed with an exchange-correlation functional within the generalized gradient approximation (GGA) which allow for vertical displacement to minimize the total free energy. The energy gain $\Delta\Omega_{f}^{GGA}$~=-0.7 eV per unit cell compared to the bare Ge(001) surface was already reported in Ref. \cite{PhysRevB.81.075412}, see the equivalent \textit{AD/HD} model therein. The STM overlay of the top- and side-view (occupied states, -1 V) in Fig.~\ref{fig3}(a) and (b) \textit{qualitatively} resembles the experiment of Fig.\ref{fig2}(a), albeit with a reduced height of the wires and a more pronounced zig-zag along the chain direction. Most notably the distinct charge clouds along the wire result from Ge not from Au. Good agreement exists for the structureless trenches where the non-buckled Au dimers form a ladder arrangement with a 4 \AA\ periodicity of low contrast in the calculated STM images. 

For a second cross-check the Patterson map for the minimum model is calculated, see Fig.\ref{fig3}(c). The non-relaxed configuration (red) yields the same positions for the maxima as in the experiment with minor deviations for the intensity modulation. Here,  $|\vec{d}_2|$ is equal to $|\vec{d}_3|$ and $|\vec{d}_1|$ has the same intensity as $|\vec{d}_4|$. Going now to the relaxed coordinates from DFT yields a totally different Patterson map (black) with a splitting of maxima related to $|\vec{d}_{2/3/4}|$, see overlay with experiment in Fig.\ref{fig3}(d). One origin of these discrepancies must be the buckling of the hetero-dimer caused by the DFT relaxation, see side view in Fig.~\ref{fig3}(a) where the Ge atom is slightly located above the Au atom. Consequently, this structural model already containing some approximations is not sufficient after DFT relaxation to account for all of the experimental findings. Thus, a more refined approach may be needed based on more extensive data.  

In summary, the in-plane data from SXRD combined with the accurate Au coverage yield the essential distances to model Au atoms in the c(8$\times$2) unit cell of Au/Ge(001). Simple structural building blocks are Au-homo- and Au-Ge hetero-dimers. The former is compatible with the trenches from both STM and DFT topography. The latter is supported by DFT as the wire building block, where Ge orbitals are the main contribution to the nanowire topography in STM.  A cross-check for any structural model is provided by the calculated Patterson map, which in the present case yields some discrepancies for the hetero-dimer after relaxation in DFT. Thus, additional investigations are highly desirable, e.g. the complete determination of the SXRD crystal truncation rods, to experimentally account for \textit{vertical} relaxation. Further attempts may also try to address the weak p(4$\times$1) superstructure.

\begin{acknowledgments}
The authors thank L.~Patthey, M.~Shi and X.Y.~Cui for making sample preparation available 
at the SIS beamline X09LA, Swiss Light Source. Technical support by M.~Lange is gratefully acknowledged. Financial
support is granted by DFG (FOR 1162 and Scha 1510/2-1) and the European Community's Seventh Framework Programme (FP7/2007-2013) under grant agreement no.$^\circ$226716 (for ELISA).
\end{acknowledgments}

\end{document}